\documentclass[conference]{IEEEtran}
\IEEEoverridecommandlockouts
\usepackage{amsmath}
\usepackage{amssymb}
\usepackage{mathtools}
\usepackage{amsfonts}
\usepackage{graphicx}
\usepackage{epsfig}
\usepackage{subfigure}
\usepackage{psfrag}
\usepackage{xcolor}
\usepackage{cite}
\usepackage{url}
\usepackage[colorlinks,linkcolor=black,urlcolor=black,anchorcolor=black,citecolor=black,hyperfootnotes=true]{hyperref}
\def\BibTeX{{\rm B\kern-.05em{\sc i\kern-.025em b}\kern-.08em
    T\kern-.1667em\lower.7ex\hbox{E}\kern-.125emX}}

\newtheorem{theorem}{Theorem}

\newcommand{\mv}[1]{\mbox{\boldmath{$ #1 $}}}
\begin{document}

\title{Channel Estimation for Intelligent Reflecting Surface Assisted Multiuser Communications}

\author{\IEEEauthorblockN{Zhaorui Wang${}^\dagger$, Liang Liu${}^\dagger$, and Shuguang Cui${}^\ddagger{}^\ast$}
	\IEEEauthorblockA{${}^\dagger$ EIE Department, The Hong Kong Polytechnic University. Email: \{zhaorui.wang,liang-eie.liu\}@polyu.edu.hk\\
${}^\ddagger$ The Chinese University of Hong Kong, Shenzhen and SRIBD. Email: shuguangcui@cuhk.edu.cn\\
${}^\ast$ ECE Department, University of California, Davis.}
}

\maketitle

\begin{abstract}
In the intelligent reflecting surface (IRS) assisted communication systems, the acquisition of channel state information (CSI) is a crucial impediment for achieving the passive beamforming gain of IRS because of the considerable overhead required for channel estimation. Specifically, under the current beamforming design for IRS-assisted communications, $KMN+KM$ channel coefficients should be estimated if the passive IRS cannot estimate its channels with the base station (BS) and users due to its lack of radio frequency (RF) chains, where $K$, $N$ and $M$ denote the numbers of users, reflecting elements of the IRS, and antennas at the BS, respectively. These numbers can be extremely large in practice considering the current trend of massive MIMO (multiple-input multiple-output), i.e., a large $M$, and massive connectivity, i.e., a large $K$. To accurately estimate such a large number of channel coefficients within a short time interval, we devote our endeavour in this paper to investigating the efficient pilot-based channel estimation method in IRS-assisted uplink communications. Building upon the observation that each IRS element reflects the signals from all the users to the BS via the same channel, we analytically verify that a time duration consisting of $K+N+\max(K-1,\lceil (K-1)N/M \rceil)$ pilot symbols is sufficient for the BS to perfectly recover all the $KMN+KM$ channel coefficients for the case without receiver noise. In contrast to the conventional uplink communications without IRS in which the minimum pilot sequence length is independent with the number of receive antennas, our study reveals the significant role of massive MIMO in reducing the channel training time for IRS-assisted communications.
\end{abstract}
\vspace{-4pt}
\section{Introduction}\label{sec:Introduction}
\vspace{-4pt}

\begin{figure}[t]
  \centering
  \includegraphics[width=8.5cm]{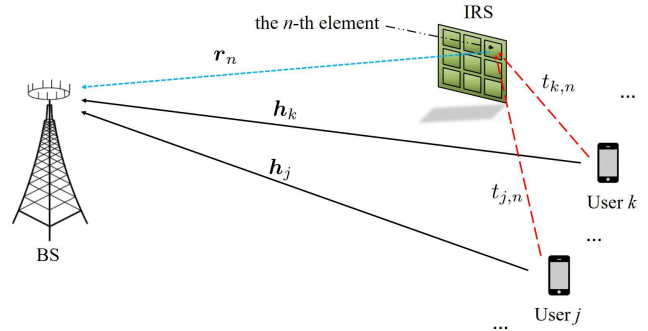}\vspace{-10pt}
  \caption{An IRS-assisted multiuser communication system.}\label{Fig1}\vspace{-20pt}
\end{figure}

Recently, intelligent reflecting surface (IRS) and its various equivalents have emerged as a promising solution to enhance the network throughput \cite{Liaskos08,Renzo19,Basar19}, thanks to their capability of modifying the wireless channels between the base station (BS) and users to be more favorable for their communications via inducing an independent phase shift to the incident signal at each reflecting element in real-time, as shown in Fig. \ref{Fig1}. Assuming perfect channel state information (CSI) at the BS, the joint beamforming optimization at the BS and IRS has been studied under various setups (see, e.g., \cite{Wu18,Zhang19}), which shows the effectiveness of IRS in enhancing the system throughput.

However, the above throughput gain in the IRS-assisted communication systems critically depends on channel estimation for acquiring CSI, which is quite challenging in practice. Specifically, to reduce the implementation cost, the IRS generally is not equipped with any radio frequency (RF) chain and thus lacks the baseband processing capability. As a result, it is impossible for the IRS to estimate its channels with the BS and users. In this case, in a single cell consisting of a BS with $M$ antennas, $K$ single-antenna users, and one IRS with $N$ reflecting elements, the algorithms for designing the beamforming vectors used by the BS and IRS \cite{Wu18,Zhang19} require the system to estimate $KMN+KM$ channel coefficients, which can be extremely large considering the current trend towards massive multiple-input multiple-output (MIMO) \cite{Larsson14} and massive connectivity \cite{Liu18}. Recently, several works have proposed various strategies to efficiently estimate the channels in the IRS-assisted single-user communication systems \cite{Mishra19,Yang19,Zheng19,Jensen}. However, to our best knowledge, channel estimation for IRS-assisted multiuser communications still remains an open problem due to the significantly larger number of involved channel coefficients compared to the single-user counterpart.

Motivated by the above challenge, this work aims to characterize the minimum pilot sequence length required to estimate the CSI in the IRS-assisted multiuser uplink communications for the first time in the literature. Specifically, we consider the passive pilot based channel estimation approach, in which the IRS elements passively reflect the pilot sequences sent by the users to the BS such that it is able to estimate the CSI associated with the IRS. By leveraging the observation that each IRS element reflects the signals from all the users to the BS via the same channel, as shown in Fig. \ref{Fig1}, our main result is that perfect channel estimation for the case without noise at the BS can be achieved over a time duration merely consisting of $K+N+\max(K-1,\lceil(K-1)N/M \rceil)$ pilot symbols. The achievability is validated by an innovative \emph{three-phase channel estimation strategy}. In Phase I consisting of $K$ symbols, the IRS is switched off such that the BS can estimate its direct channels with the users. In Phase II with $N$ symbols, only one typical user is selected to transmit the non-zero pilot symbols and the IRS reflected channel for this typical user is estimated. In Phase III of $\max(K-1,\lceil(K-1)N/M \rceil)$ symbols, the IRS reflected channels of the other users are estimated based on the fact that these channels are scaled versions of the typical user's reflected channels. Different from the multiple-access channel where the minimum pilot sequence length has nothing to do with the number of receive antennas \cite{Hassibi03}, our strategy can make the best of the receive antennas to reduce the channel training duration in IRS-assisted multiuser systems. Especially, in the massive MIMO regime with $M>N$, the minimum pilot sequence length under our strategy is $2K+N-1$, which is scalable with the number of users: if there is one more user in the system, only 2 additional pilot symbols are sufficient to estimate the new $MN+M$ channel coefficients associated with this user.

\vspace{-2pt}
\section{System Model}\label{sec:SYS}
\vspace{-2pt}

We consider a narrow-band wireless system with $K$ single-antenna users simultaneously communicating with a BS equipped with $M$ antennas in the uplink, where an IRS equipped with $N$ passive reflecting elements is deployed to enhance the communication performance, as shown in Fig. \ref{Fig1}. We assume quasi-static block-fading channels, in which all the channels remain approximately constant in each fading block with $T$ symbols. Let $\mv{h}_k\in\mathbb{C}^{M\times1}$, $k=1,\cdots,K$, denote the direct channel from the $k$th user to the BS. Further, let $t_{k,n}\in\mathbb{C}$ and $\mv{r}_n\in\mathbb{C}^{M\times1}$ denote the channels from the $k$th user to the $n$th IRS element and from the $n$th IRS element to the BS, respectively, $k=1,\cdots,K$, $n=1,\dots,N$. Moreover, $\mv{h}_k$'s, $t_{k,n}$'s, and $\mv{r}_n$'s are all assumed to follow the independent and identically distributed (i.i.d.) Rayleigh fading channel model, i.e., $\mv{h}_k \sim \mathcal{CN}(0,\beta_k^{{\rm BU}}\mv{I})$, $t_{k,n}\sim \mathcal{CN}(0,\beta_{k,n}^{{\rm IU}})$, and $\mv{r}_n\sim \mathcal{CN}(\mv{0},\beta_n^{{\rm BI}}\mv{I})$, $\forall k,n$, where $\beta_k^{{\rm BU}}$, $\beta_{k,n}^{{\rm IU}}$, and $\beta_n^{{\rm BI}}$ denote the path loss of $\mv{h}_k$, $t_{k,n}$ and $\mv{r}_n$, respectively.

Thanks to the IRS controller, each element on IRS is able to dynamically adjust its reflection coefficient for re-scattering the electromagnetic waves from the users to the BS such that the useful signal and harmful interference can be added at the BS in constructive and destructive manners, respectively \cite{Liaskos08,Renzo19,Basar19}. Specifically, let $\phi_{n,i}$ denote the reflection coefficient of the $n$th IRS element at the $i$th time instant, $n=1,\cdots,N$, $i=1,\cdots,T$, which satisfies
\begin{equation}
|\phi_{n,i}|=\left\{\begin{array} {ll} 1, ~ {\rm if ~ element} ~ n ~ {\rm is ~ on ~ at ~ time ~ instant} ~ i, \\ 0, ~ {\rm otherwise}. \end{array} \right. \label{eq:Sys-1}
\end{equation}In other words, if an IRS element is switched on, it can only change the phase of its incident signal \cite{Wu18,Zhang19}.

With the existence of the IRS, the received signal of the BS at time instant $i$, $i=1,\cdots,T$, which is a superposition of the signals from the users' direct communication links and the reflected links via the IRS, is expressed as
\begin{align}
\mv{y}^{(i)}&=\sum_{k=1}^{K}\mv{h}_k\sqrt{p}x_k^{(i)}+\sum_{k=1}^{K}\sum_{n=1}^{N}\phi_{n,i}t_{k,n}\mv{r}_n\sqrt{p}x_k^{(i)}+\mv{z}^{(i)} \nonumber \\
&=\sum_{k=1}^{K}\left(\mv{h}_k+\sum_{n=1}^{N}\phi_{n,i}\mv{g}_{k,n}\right)\sqrt{p}x_k^{(i)}+\mv{z}^{(i)},
\label{eq:Sys-1.5}
\end{align}where $x_k^{(i)}$ and $\mv{z}^{(i)}\sim\mathcal{CN}\left(\mv{0},\sigma^2\mv{I}\right)$ denote the transmit signal of user $k$ and additive white Gaussian noise (AWGN) at the BS at time instant $i$, respectively, $p$ denotes the identical transmit power of the users, and\vspace{-6pt}
\begin{align}\label{eqn:effective channel}
\mv{g}_{k,n}=t_{k,n}\mv{r}_n, ~~~ \forall n,k, \vspace{-4pt}
\end{align}denotes the effective channel from the $k$th user to the BS through the $n$th IRS element.

In this paper, we consider the legacy two-stage transmission protocol for the uplink communications, in which each coherence block of length $T$ symbols is divided into the channel estimation stage consisting of $\tau<T$ symbols and data transmission stage consisting of $T-\tau$ symbols. Specifically, in the channel estimation stage, each user $k$ is assigned with a pilot sequence consisting of $\tau$ symbols:
\begin{align}\label{eqn:pilot channel}\vspace{-6pt}
\mv{a}_k=[a_{k,1},\cdots,a_{k,\tau}]^T, ~~~ k=1,\cdots,K, \vspace{-6pt}
\end{align}where the norm of $a_{k,i}$ is either zero or one, $\forall k,i$. At time instant $i\leq \tau$, user $k$ transmits $x_{k,i}=a_{k,i}$ to the BS, and the received signal at the BS is
\begin{align}\label{eqn:received pilot}
\mv{y}^{(i)}=\sum_{k=1}^{K}\left(\mv{h}_k+\sum_{n=1}^{N}\phi_{n,i}\mv{g}_{k,n}\right)\sqrt{p}a_{k,i}&+\mv{z}^{(i)}, i\leq \tau.
\end{align}Define $\mv{Y}=[\mv{y}^{(1)},\cdots,\mv{y}^{(\tau)}]\in \mathbb{C}^{M\times \tau}$ as the overall received signal at the BS across all the $\tau$ time instants of the channel estimation phase. The task of the BS is then to estimate all the direct channels and reflected channels based on the received signal $\mv{Y}$ as well as its knowledge of the known user pilot symbols $a_{k,i}$'s and IRS reflection coefficients $\phi_{n,i}$'s.


In the data transmission stage, the reflection coefficient of each IRS element $n$ is fixed over different time instants, i.e., $\phi_{n,i}=\phi_n$, $i=\tau+1,\cdots,T$ \cite{Wu18,Zhang19}. Moreover, to convey the information, the transmit symbol of user $k$ in the $i$th time instant is modeled as a circularly symmetric complex Gaussian (CSCG) random variable with zero mean and unit variance, i.e., $x_k^{(i)}\sim \mathcal{CN}(0,1)$, $k=1,\cdots,K$, $i=\tau+1,\cdots,T$. Then, at each time instant $i=\tau+1,\cdots,T$, the BS applies the beamforming vector $\mv{w}_k$ to decode the message of user $k$, i.e.,
\begin{align}
\mv{\tilde{y}}_k^{(i)}=\sum_{q=1}^K\mv{w}_k^H\left(\mv{h}_q\hspace{-2pt}+\hspace{-2pt}\sum_{n=1}^{N}\phi_n\mv{g}_{q,n}\right)\sqrt{p}x_q^{(i)}\hspace{-2pt}+\hspace{-2pt}\mv{w}_k^H\mv{z}^{(i)}.
\end{align}Therefore, the achievable rate of user $k$, $k=1,\cdots,K$, is
\begin{align}
R_k=\frac{T-\tau}{T}\log_2(1+\gamma_k),
	\label{eq:Sys-1.6}
\end{align}where $\frac{T-\tau}{T}$ denotes the fraction of time for data transmission, and
\begin{align}
\gamma_k=\frac{p\left|\mv{w}_k^H\left(\mv{h}_k\hspace{-2pt}+\hspace{-2pt}\sum\limits_{n=1}^{N}\phi_n\mv{g}_{k,n}\right)\right|^2}{\sum\limits_{q\ne k}p\left|\mv{w}_k^H\left(\mv{h}_q\hspace{-2pt}+\hspace{-2pt}\sum\limits_{n=1}^{N}\phi_n\mv{g}_{q,n}\right)\right|^2\hspace{-2pt}+\hspace{-2pt}\sigma^2\|\mv{w}_k\|^2}.
\end{align}

It is observed from (\ref{eq:Sys-1.6}) that to improve the user rate by jointly optimizing the receive beamforming vectors $\mv{w}_k$'s at the BS and reflection coefficients $\phi_n$'s at the IRS, the BS has to know all the $MK+MKN$ channel coefficients in $\mv{h}_k$'s, $k=1,\cdots,K$, and $\mv{g}_{k,n}$'s, $k=1,\cdots,K$, $n=1,\cdots,N$. Note that $MK+MKN$ is generally a very large number in next-generation cellular networks with large-scale antenna arrays at the BSs (i.e., large $M$) and a massive number of connecting users (i.e., large $K$). Further, to increase the time duration for data transmission in (\ref{eq:Sys-1.6}), very few pilot symbols can be utilized for estimating such a large number of channel coefficients. To tackle the above issues, in the rest of this paper, we mainly focus on the channel training stage in our considered system, and propose an innovative scheme to estimate $\mv{h}_k$'s and $\mv{g}_{k,n}$'s accurately with low training overhead.

\section{Three-Phase Channel Estimation Protocol}\label{sec:channel estimation model}
In this section, we propose a novel three-phase channel estimation protocol for IRS-assisted multiuser communications. The main idea is that although $KMN$ unknowns need to be estimated in $\mv{g}_{k,n}$'s, the degrees-of-freedom (DoF) for all these channel coefficients is much smaller than $KMN$. Specifically, each $\mv{r}_n$ appears in all $\mv{g}_{k,n}$'s, $\forall k$, according to (\ref{eqn:effective channel}), since each IRS element reflects the signals from all the $K$ users to the BS via the same channel. It is thus theoretically feasible to employ much fewer pilot symbols to estimate the $KMN$ unknowns in $\mv{g}_{k,n}$'s via leveraging their strong correlations. Nevertheless, it is challenging to exploit the correlations among the channel coefficients arising from $\mv{r}_n$'s, since the IRS cannot estimate $\mv{r}_n$'s due to the lack of RF chains.

\begin{figure}[t]
  \centering
  \includegraphics[width=9cm]{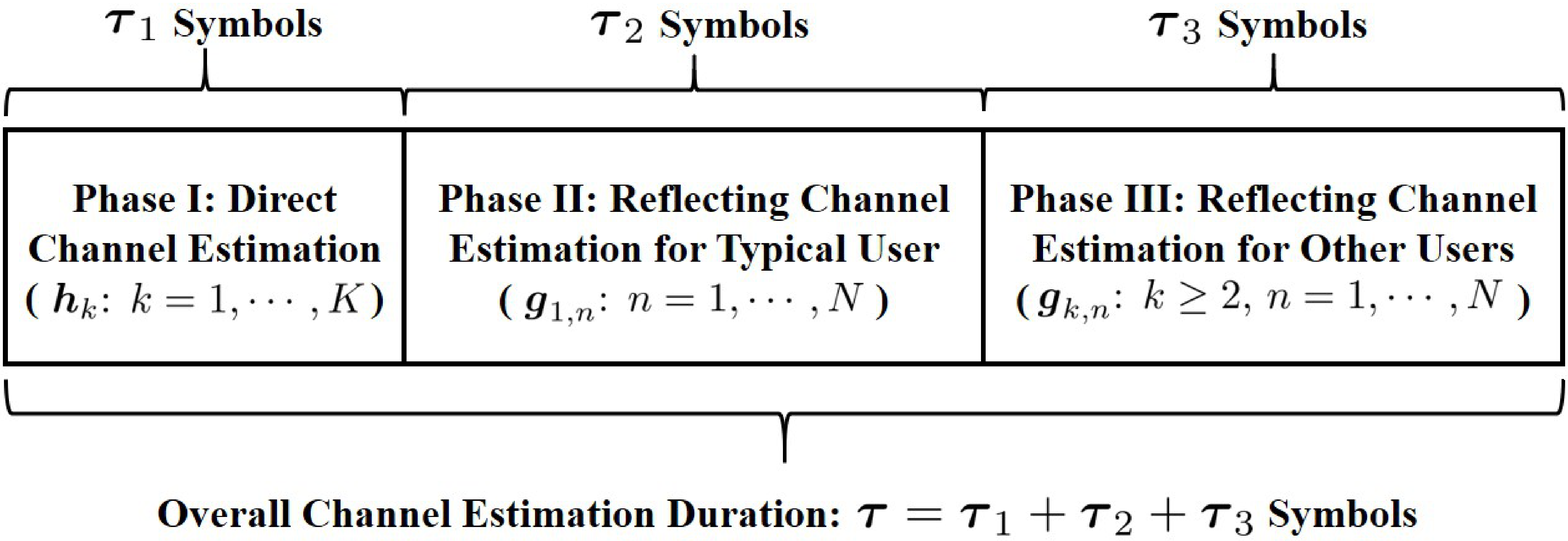}\vspace{-5pt}
  \caption{The proposed three-phase channel estimation protocol.}\label{Fig2}\vspace{-10pt}
\end{figure}

In the following, we introduce our proposed three-phase channel estimation protocol to tackle the above issue. As shown in Fig. \ref{Fig2}, in the first phase consisting of $\tau_1$ symbols, the BS estimates the direct channels $\mv{h}_k$'s by switching off all IRS elements; in the second phase consisting of $\tau_2$ symbols, merely one \emph{typical user}, denoted by user $1$ for convenience, transmits non-zero pilot symbols, based on which the BS estimates the reflected channels of this typical user, i.e., $\mv{g}_{1,n}$'s, $\forall n$, with all IRS element switched on; in the third phase consisting of $\tau_3=\tau-\tau_1-\tau_2$ symbols, the BS estimates the reflected channels of the other users based on their following relationship with the reflected channels of user 1:
\begin{align}\label{eqn:correlated channel}
\mv{g}_{k,n}=\lambda_{k,n}\mv{g}_{1,n}, ~~~ k=2,\cdots,K, ~ n=1,\cdots,N,
\end{align}where
\begin{align}\label{eqn:alpha}
\lambda_{k,n}=\frac{t_{k,n}}{t_{1,n}}, ~~~ k=2,\cdots,K, ~ n=1,\cdots,N.
\end{align}With known $\mv{g}_{1,n}$'s, $n=1,\cdots,N$, (\ref{eqn:correlated channel}) reveals that each channel vector $\mv{g}_{k,n}$ with $k\geq 2$ can be efficiently recovered via merely estimating a scalar $\lambda_{k,n}$. As a result, the channel training time in this phase can be significantly reduced.

For the purpose of drawing essential insights, in the rest of this paper, we mainly focus on the ideal case without noise at the BS, i.e., $\mv{z}^{(i)}=\mv{0}$, $\forall i$, and characterize the \emph{minimum pilot sequence length} to estimate all the channels perfectly under the proposed protocol. Such a result can theoretically demonstrate the performance gain brought by this new protocol for channel estimation in our considered system. The issue of how to utilize our proposed strategy to estimate the channels for the practical case with noise at the BS will be left to our future works.

\section{Performance Limits for the Case without Receiver Noise}
In this section, we consider the ideal case without noise at the BS. In this scenario, the proposed three-phase channel estimation protocol works as follows.

\vspace{-2pt}
\subsection{Phase I: Direct Channel Estimation}\vspace{-2pt}
To estimate $\mv{h}_k$'s in Phase I, the IRS is switched off, i.e.,
\begin{align}\label{eqn:off}
\phi_{n,i}=0, ~~~ n=1,\cdots,N, ~ i=1,\cdots,\tau_1.
\end{align}According to \cite{Hassibi03}, each user $k$ can send
\begin{align}\label{eqn:length of phase 1}
\tau_1\geq \tilde{\tau}_1=K,
\end{align}pilot symbols, denoted by $\mv{a}_k^{{\rm I}}=[a_{k,1},\cdots,a_{k,\tau_1}]^T$, $\forall k$, to the BS for channel estimation. Then, based on (\ref{eqn:received pilot}), the received signal at the BS over the whole phase is
\begin{align}\label{eqn:received pilot direct channel}
\mv{Y}^{{\rm I}}\hspace{-2pt}=\hspace{-2pt}[\mv{y}^{(1)},\cdots,\mv{y}^{(\tau_1)}] =\sqrt{p}[\mv{h}_1,\cdots,\mv{h}_K]\left[\hspace{-2pt}\begin{array}{c}(\mv{a}_1^{{\rm I}})^T \\ \vdots \\ (\mv{a}_K^{{\rm I}})^T\end{array}\hspace{-2pt}\right].
\end{align}As a result, the direct channels $\mv{h}_k$'s can be estimated perfectly by solving (\ref{eqn:received pilot direct channel}) if the pilot sequences of different users are orthogonal to each other, i.e.,
\begin{align}\label{eqn:full rank}
\left[\begin{array}{c}(\mv{a}_1^{{\rm I}})^T \\ \vdots \\ (\mv{a}_K^{{\rm I}})^T\end{array}\right][(\mv{a}_1^{{\rm I}})^\ast,\cdots,(\mv{a}_K^{{\rm I}})^\ast]=\tau_1\mv{I}.
\end{align}Since each pilot sequence consists of $\tau_1\geq K$ symbols, it is feasible to design $\mv{a}_k^{{\rm I}}$'s to satisfy (\ref{eqn:full rank}). Then, according to (\ref{eqn:received pilot direct channel}), $\mv{h}_k$'s can be perfected estimated as
\begin{align}\label{eqn:perfect channel estimation phase 1}
[\mv{h}_1,\cdots,\mv{h}_K]=\frac{1}{\tau_1\sqrt{p}}\mv{Y}^{{\rm I}}[(\mv{a}_1^{{\rm I}})^\ast,\cdots,(\mv{a}_K^{{\rm I}})^\ast].
\end{align}

\begin{figure*}[ht]
\setcounter{equation}{27}
\begin{align}\label{eqn:pilot array}
& \mv{V}=\left[\begin{array}{ccccccc} \phi_{1,\theta+1}a_{2,\theta+1}\mv{g}_{1,1} & \cdots & \phi_{N,\theta+1}a_{2,\theta+1}\mv{g}_{1,N} & \cdots & \phi_{1,\theta+1}a_{K,\theta+1}\mv{g}_{1,1} & \cdots & \phi_{N,\theta+1}a_{K,\theta+1}\mv{g}_{1,N} \\ \vdots & \ddots & \vdots & \ddots & \vdots & \ddots & \vdots \\ \phi_{1,\theta+\tau_3}a_{2,\theta+\tau_3}\mv{g}_{1,1} & \cdots & \phi_{N,\theta+\tau_3}a_{2,\theta+\tau_3}\mv{g}_{1,N} & \cdots & \phi_{1,\theta+\tau_3}a_{K,\theta+\tau_3}\mv{g}_{1,1} & \cdots & \phi_{N,\theta+\tau_3}a_{K,\theta+\tau_3}\mv{g}_{1,N} \end{array} \right], \\
& {\rm with} ~ \theta=\tau_1+\tau_2. \nonumber
\end{align}  \setcounter{equation}{15}
\hrulefill
\end{figure*}

\vspace{-2pt}
\subsection{Phase II: Reflecting Channel Estimation for Typical User}\vspace{-2pt}
In the second phase, the IRS is switched on. Define $\mv{a}_k^{{\rm II}}=[a_{k,\tau_1+1},\cdots,a_{k,\tau_1+\tau_2}]^T$ as the pilot sequence of user $k$ in Phase II, $\forall k$. To estimate $\mv{g}_{1,n}$'s, we set
\begin{align}\label{eqn:zero pilot phase 2}
\mv{a}_k^{{\rm II}}=\mv{0},  ~~~ k=2,\cdots,K.
\end{align}Note that $\mv{h}_k$'s have already been perfectly estimated by (\ref{eqn:perfect channel estimation phase 1}) in Phase I. Therefore, their interference for estimating $\mv{g}_{1,n}$'s can be canceled from the received signal at the BS in Phase II. According to (\ref{eqn:received pilot}), after interference cancellation, the effective received signal at the BS at time instant $i$ in Phase II is
\begin{align}\label{eqn:received pilot reflected channel user 1}
\bar{\mv{y}}^{(i)}&=\mv{y}^{(i)}-\sum_{k=1}^{K}\mv{h}_k\sqrt{p}a_{k,i} \nonumber \\ &= \sum_{n=1}^{N}\phi_{n,i}\mv{g}_{1,n}\sqrt{p}a_{1,i}, ~ i=\tau_1+1,\cdots,\tau_1+\tau_2.
\end{align}The overall effective received signal at the BS in the second phase is then expressed as
\begin{align}\label{eqn:received pilot phase 2}
\hspace{-4pt}\bar{\mv{Y}}^{{\rm II}}&\hspace{-2pt}=\hspace{-2pt}[\bar{\mv{y}}^{(\tau_1+1)},\cdots,\bar{\mv{y}}^{(\tau_1+\tau_2)}] \nonumber \\
&\hspace{-2pt}=\hspace{-2pt} \sqrt{p}[\mv{g}_{1,1},\hspace{-1pt}\cdots \hspace{-1pt},\mv{g}_{1,N}]\mv{\Phi}^{{\rm II}}{\rm diag}(\mv{a}_1^{{\rm II}}),
\end{align}where
\begin{align}\label{eqn:Phi}
\mv{\Phi}^{{\rm II}}=\left[\begin{array}{ccc}\phi_{1,\tau_1+1} & \cdots & \phi_{1,\tau_1+\tau_2} \\ \vdots & \ddots & \vdots \\ \phi_{N,\tau_1+1} & \cdots & \phi_{N,\tau_1+\tau_2} \end{array} \right].
\end{align}To solve (\ref{eqn:received pilot phase 2}), we can simply set
\begin{align}\label{eqn:pilot phase 2}
a_{1,i}=1, ~~~ i=\tau_1+1,\cdots,\tau_1+\tau_2.
\end{align}In this case, it can be shown that as long as
\begin{align}\label{eqn:tau 2}
\tau_2\geq \tilde{\tau}_2=N,
\end{align}we can always construct a $\mv{\Phi}^{{\rm II}}$ such that ${\rm rank}(\mv{\Phi}^{{\rm II}})=N$ under the constraint given in (\ref{eq:Sys-1}). The construction of such a $\mv{\Phi}^{{\rm II}}$ can be based on the discrete Fourier transform (DFT) matrix
\begin{align}\label{eqn:DFT}
\hspace{-4pt} \mv{\Phi}^{{\rm II}}=\left[\begin{array}{ccccc} 1 & 1 & 1 & \cdots & 1 \\ 1 & \omega & \omega^2 & \cdots & \omega^{\tau_2-1} \\ 1 & \omega^2 & \omega^4 & \cdots & \omega^{2(\tau_2-1)} \\ \vdots & \vdots & \vdots & \ddots & \vdots \\ 1 & \omega^{N-1} & \omega^{2(N-1)} & \cdots & \omega^{(N-1)(\tau_2-1)} \end{array} \right],
\end{align}where $\omega=e^{-2\pi j/\tau_2}$ with $j^2=-1$. In this case, it follows that $\mv{\Phi}^{{\rm II}}(\mv{\Phi}^{{\rm II}})^H=\tau_2\mv{I}$. As a result, the reflected channels of user $1$ can be perfectly estimated as
\begin{align}\label{eqn:perfect channel estimation phase 2}
[\mv{g}_{1,1},\cdots,\mv{g}_{1,N}]=\frac{1}{\tau_2\sqrt{p}}\bar{\mv{Y}}^{{\rm II}}(\mv{\Phi}^{{\rm II}})^H.
\end{align}

\subsection{Phase III: Reflecting Channel Estimation for Other Users}
In Phase III, let $\mv{a}_k^{{\rm III}}=[a_{k,\tau_1+\tau_2+1},\cdots,a_{k,\tau_1+\tau_2+\tau_3}]^T$ denote the pilot sequence of user $k$, $\forall k$. To estimate $\mv{g}_{k,n}$'s, $\forall k\geq 2$, $\forall n$, we set
\begin{align}\label{eqn:zero pilot phase 3}
\mv{a}_1^{{\rm III}}=\mv{0}.
\end{align}One straightforward manner is to allow only one user $k\geq 2$ to transmit $\tau_2 \geq N$ pilot symbols each time such that its reflected channels $\mv{g}_{k,n}$'s, $\forall n$, can be directly estimated based on the approach for estimating $\mv{g}_{1,n}$'s. Under such a scheme, at least we need to use $\tau_3=(K-1)N$ time instants in total to estimate the reflecting channels of the remaining $K-1$ users. However, with a large number of users, the estimation of $\mv{g}_{k,n}$'s will take quite a long time, which leads to reduced user transmission rate due to the limited time left for data transmission as shown in (\ref{eq:Sys-1.6}).

We propose to exploit the channel correlations among $\mv{g}_{k,n}$'s to reduce the channel estimation time in Phase III. Specifically, similar to (\ref{eqn:received pilot reflected channel user 1}), after cancelling the interference caused by the direct channels $\mv{h}_k$'s, the effective received signal at the BS at each time instant $i$, $i=\tau_1+\tau_2+1,\cdots,\tau_1+\tau_2+\tau_3$, in Phase III is
\begin{align}
\bar{\mv{y}}^{(i)}&=\mv{y}^{(i)}-\sum_{k=1}^{K}\mv{h}_k\sqrt{p}a_{k,i} \nonumber \\ &= \sum\limits_{k=2}^K\sum_{n=1}^{N}\phi_{n,i}\mv{g}_{k,n}\sqrt{p}a_{k,i}, \label{eqn:received pilot reflected channel user k 1}  \\
& = \sum\limits_{k=2}^K\sum_{n=1}^{N}\phi_{n,i}\lambda_{k,n}\mv{g}_{1,n}\sqrt{p}a_{k,i}. \label{eqn:received pilot reflected channel user k 2}
\end{align}The overall effective received signal at the BS in Phase III is
\begin{align}\label{eqn:received pilot phase 3}
\hspace{-6pt}\bar{\mv{y}}^{{\rm III}}\hspace{-2pt}=\hspace{-2pt}\left[\left(\bar{\mv{y}}^{(\tau_1+\tau_2+1)}\right)^T,\cdots,\left(\bar{\mv{y}}^{(\tau_1+\tau_2+\tau_3)}\right)^T\right]^T\hspace{-2pt}=\hspace{-2pt}\sqrt{p}\mv{V}\mv{\lambda},
\end{align}where $\mv{\lambda}=[\mv{\lambda}_2^T,\cdots,\mv{\lambda}_K^T]^T\in \mathbb{C}^{(K-1)N\times 1}$ with $\mv{\lambda}_k=[\lambda_{k,1} \cdots \lambda_{k,N}]^T$, $k=2,\cdots,K$, and $\mv{V}\in \mathbb{C}^{M\tau_3\times (K-1)N}$ is given in (\ref{eqn:pilot array}) on the top of this page.

Mathematically, (\ref{eqn:received pilot phase 3}) defines an equivalent linear channel estimation model consisting of $(K-1)N$ users, where each column of $\mv{V}$ denotes the pilot sequence sent by each of these effective users. One interesting observation of $\mv{V}$ here is that thanks to the multiple antennas at the BS, the effective channel estimation time is increased from $\tau_3$ to $M\tau_3$. In other words, it is possible to adopt the multi-antenna technology to significantly reduce the channel training time in Phase III under our proposed strategy.

In the following, via a proper design of $\mv{a}_k^{{\rm III}}=[a_{k,\tau_1+\tau_2+1},\cdots,a_{k,\tau_1+\tau_2+\tau_3}]^T$'s, $k=2,\cdots,K$, and $\phi_{n,i}$'s, $n=1,\cdots,N$, $i=\tau_1+\tau_2+1,\cdots,\tau_1+\tau_2+\tau_3$, we aim to find the minimum value of $\tau_3$ to satisfy ${\rm rank}(\mv{V})=(K-1)N$ such that $\mv{\lambda}$ can be perfectly estimated based on (\ref{eqn:received pilot phase 3}). We start with the case of $M\geq N$.

\begin{theorem}\label{theorem1}
In the case of $M\geq N$, the minimum value of $\tau_3$ to guarantee perfect estimation of $\mv{\lambda}$ according to (\ref{eqn:received pilot phase 3}) is given by
\setcounter{equation}{28}\begin{align}\label{eqn:tau 3}
\tilde{\tau}_3=K-1.
\end{align}To achieve perfect estimation of $\mv{\lambda}$ given the above minimum value of $\tau_3$, we can set
\begin{align}
& \hspace{-2pt} a_{k,i}=\left\{\begin{array}{ll}1, & {\rm if} ~ k-1=i-\tau_1-\tau_2, \\ 0, & {\rm otherwise}, \end{array}\right. \nonumber \\ & ~~~~~~~~~~~ 2\leq k \leq K, ~ \tau_1\hspace{-2pt}+\hspace{-2pt}\tau_2\hspace{-2pt}+\hspace{-2pt}1\leq i \leq \tau_1\hspace{-2pt}+\hspace{-2pt}\tau_2\hspace{-2pt}+\hspace{-2pt}K\hspace{-2pt}-\hspace{-2pt}1, \label{eqn:a case 1}  \\
& \hspace{-2pt} \phi_{n,i}=1, ~ 1\leq n \leq N, ~ \tau_1\hspace{-2pt}+\hspace{-2pt}\tau_2\hspace{-2pt}+\hspace{-2pt}1\leq i \leq \tau_1\hspace{-2pt}+\hspace{-2pt}\tau_2\hspace{-2pt}+\hspace{-2pt}K\hspace{-2pt}-\hspace{-2pt}1. \label{eqn:b case 1}
\end{align}Then, $\mv{\lambda}$ can be perfectly estimated as
\begin{align}\label{eqn:estimation 1}
\hspace{-2pt}\mv{\lambda}_k=[\mv{g}_{1,1},\cdots,\mv{g}_{1,N}]^{\dag}\frac{\bar{\mv{y}}^{(\tau_1+\tau_2+k-1)}}{\sqrt{p}}, ~ k=2,\cdots,K,
\end{align}where for any matrix $\mv{B}\in \mathbb{C}^{x\times y}$ with $x\geq y$, $\mv{B}^\dag=(\mv{B}^H\mv{B})^{-1}\mv{B}^H$ denotes its pseudo-inverse matrix.
\end{theorem}

\begin{IEEEproof}
Please refer to Appendix \ref{appendix1}.
\end{IEEEproof}


Next, we consider the case of $M<N$. In this case, define $\rho=\lfloor\frac{N}{M}\rfloor$, $\upsilon=N-M\rho$, and $\mathcal{N}=\{1,\cdots,N\}$, where $\lfloor x \rfloor$ denotes the largest integer that is no larger than $x$. For each user $k\geq 2$, define two sets $\Lambda_{k,1}\subset \mathcal{N}$ with cardinality $|\Lambda_{k,1}|=N-\upsilon$ and $\Lambda_{k,2}\subset \mathcal{N}$ with cardinality $|\Lambda_{k,2}|=\upsilon$, which are constructed as follows. First, define
\begin{align}\label{eqn:T}
\mathcal{T}_k=\{(k-2)\upsilon +1,\cdots,(k-1)\upsilon\}, ~~~ k=2,\cdots,K.
\end{align}Then, we construct $\Lambda_{k,1}$'s and $\Lambda_{k,2}$'s as
\begin{align}
& \Lambda_{k,2}=\{m-(\lceil \frac{m}{N} \rceil-1)N:\forall m\in \mathcal{T}_k\}, \label{eqn:Lambda 2} \\
& \Lambda_{k,1}=\mathcal{N}\setminus \Lambda_{k,2}, ~~~ k=2,\cdots,K, \label{eqn:Lambda 1}
\end{align}where $\lceil x \rceil$ denotes the smallest integer that is not smaller than $x$. Moreover, for any $i=1,\cdots, (K-1)\rho$, define $\kappa_i=(i-(\lceil\frac{i}{\rho}\rceil-1)\rho-1)M$ and
\begin{align}\label{eqn:Omega}
\Omega_i=\{\Lambda_{\lceil \frac{i}{\rho}\rceil+1,1}(\kappa_i+1),\cdots,\Lambda_{\lceil \frac{i}{\rho}\rceil+1,1}(\kappa_i+M)\},
\end{align}where given any set $\mathcal{B}$, $\mathcal{B}(i)$ denotes its $i$th element. While for any $i=(K-1)\rho+1,(K-1)\rho+2,\cdots$, define
\begin{align}\label{eqn:J}
\mathcal{J}_i=&\{(i\hspace{-2pt}-\hspace{-2pt}(K\hspace{-2pt}-\hspace{-2pt}1)\rho\hspace{-2pt}-\hspace{-2pt}1)M\hspace{-2pt}+\hspace{-2pt}1,\cdots,\nonumber \\ & \min((i\hspace{-2pt}-\hspace{-2pt}(K\hspace{-2pt}-\hspace{-2pt}1)\rho)M,(K\hspace{-2pt}-\hspace{-2pt}1)N\hspace{-2pt}-\hspace{-2pt}(\hspace{-2pt}K-\hspace{-2pt}1)M\rho)\}.
\end{align}Based on $\mathcal{J}_i$, given any $i>(K-1)\rho$, we define
\begin{align}
& \mathcal{K}_i=\{\lceil\frac{j}{\upsilon}\rceil+1:\forall j \in \mathcal{J}_i\}, \label{eqn:K} \\
& \mathcal{N}_i=\{\Lambda_{\lceil \frac{j}{\upsilon}\rceil+1,2}(j-(\lceil\frac{j}{\upsilon}\rceil-1)\upsilon):\forall j\in \mathcal{J}_i\}. \label{eqn:N}
\end{align}Then, we have the following theorem.

\begin{theorem}\label{theorem2}
In the case of $M<N$, the minimum value of $\tau_3$ to guarantee perfect estimation of $\mv{\lambda}$ according to (\ref{eqn:received pilot phase 3}) is given by
\begin{align}\label{eqn:tau 3 case 2}
\tilde{\tau}_3=\left\lceil \frac{(K-1)N}{M} \right\rceil.
\end{align}To perfectly estimate $\mv{\lambda}$ given the above minimum value of $\tau_3$, at time slot $\tau_1+\tau_2+i$ with $1\leq i \leq (K-1)\rho$, we can set
\begin{align}
& \hspace{-6pt} a_{k,\tau_1+\tau_2+i}\hspace{-2pt}=\hspace{-2pt}\left\{\begin{array}{ll}\hspace{-2pt}1, & \hspace{-3pt} {\rm if} ~ k=\left \lceil \frac{i}{\rho}\right \rceil+1, \\ \hspace{-2pt}0, & \hspace{-3pt} {\rm otherwise},\end{array}\right. \label{eqn:pilot TDMA 1} \\
& \hspace{-6pt} \phi_{n,\tau_1+\tau_2+i}\hspace{-2pt}=\hspace{-2pt}\left\{\begin{array}{ll}\hspace{-2pt}1, & \hspace{-3pt} {\rm if} ~ n\in \Omega_i, \\ \hspace{-2pt} 0, & \hspace{-3pt} {\rm otherwise},\end{array}\right. ~~~ \hspace{-2pt} 1\hspace{-2pt}\leq \hspace{-2pt}i \hspace{-2pt}\leq \hspace{-2pt}(K-1)\rho. \label{eqn:reflecting coefficient TDMA}
\end{align}With the above solution, at each time instant $\tau_1+\tau_2+i$, we can perfectly estimate the following $\lambda_{k,n}$'s
\begin{align}
\hspace{-2pt} & [\lambda_{\lceil\frac{i}{\rho}\rceil+1,\Omega_i(1)},\cdots,\lambda_{\lceil\frac{i}{\rho}\rceil+1,\Omega_i(M)}]^T \nonumber \\ \hspace{-2pt} =&[\mv{g}_{1,\Omega_i(1)},\hspace{-2pt}\cdots\hspace{-2pt},\mv{g}_{1,\Omega_i(M)}]^{-1}\frac{\bar{\mv{y}}^{(\tau_1+\tau_2+i)}}{\sqrt{p}}, 1\hspace{-2pt}\leq\hspace{-2pt} i \hspace{-2pt}\leq \hspace{-2pt} (K\hspace{-2pt}-\hspace{-2pt}1)\rho. \label{eqn:perfect lambda case 1}
\end{align}Further, at time slot $\tau_1+\tau_2+i$ with $(K-1)\rho+1\leq i \leq \tilde{\tau}_3$, we can set
\begin{align}
&  a_{k,\tau_1+\tau_2+i}\hspace{-2pt}=\hspace{-2pt}\left\{\begin{array}{ll}\hspace{-2pt}1, & \hspace{-3pt} {\rm if} ~ k\in \mathcal{K}_i, \\ \hspace{-2pt}0, & \hspace{-3pt} {\rm otherwise},\end{array}\right. \label{eqn:pilot TDMA 2} \\
&  \phi_{n,\tau_1+\tau_2+i}\hspace{-2pt}=\hspace{-2pt}\left\{\begin{array}{ll}\hspace{-2pt}1, & \hspace{-3pt} {\rm if} ~ n\in \mathcal{N}_i, \\ \hspace{-2pt} 0, & \hspace{-3pt} {\rm otherwise},\end{array}\right. ~ \hspace{-2pt} (K-1)\rho+1\hspace{-2pt}\leq \hspace{-2pt}i \hspace{-2pt}\leq \hspace{-2pt}\tilde{\tau}_3. \label{eqn:reflecting coefficient TDMA 2}
\end{align}With the above solution, at each time instant $\tau_1+\tau_2+i$, we can perfectly estimate the following $\lambda_{k,n}$'s
\begin{align}
&[\lambda_{\lceil\frac{\mathcal{J}_i(1)}{\upsilon}\rceil+1,\mathcal{N}_i(1)},\cdots, \lambda_{\lceil\frac{\mathcal{J}_i(M_i)}{\upsilon}\rceil+1,\mathcal{N}_i(M_i)}]^T \nonumber \\ =&[\mv{g}_{1,\mathcal{N}_i(1)},\cdots,\mv{g}_{1,\mathcal{N}_i(M_i)}]^{\dag}\frac{\tilde{\mv{y}}^{(\tau_1+\tau_2+i)}}{\sqrt{p}}, (K\hspace{-2pt}-\hspace{-2pt}1)\rho\hspace{-2pt}+\hspace{-2pt}1\leq \hspace{-2pt} i \hspace{-2pt} \leq \hspace{-2pt} \tilde{\tau}_3, \label{eqn:perfect lambda case 2}
\end{align}where
\begin{align}
& \hspace{-5pt} M_i=|\mathcal{N}_i|, \\ & \hspace{-5pt} \tilde{\mv{y}}^{(\tau_1+\tau_2+i)}=\bar{\mv{y}}^{(\tau_1+\tau_2+i)}-\sum\limits_{k\in \mathcal{K}_i}\sum\limits_{n\in \mathcal{N}_i \cap \Lambda_{k,1}}\sqrt{p}\lambda_{k,n}\mv{g}_{1,n}. \label{eqn:signal after interference cancellation}
\end{align}
\end{theorem}

\begin{IEEEproof}
Please refer to Appendix \ref{appendix2}.
\end{IEEEproof}

According to Theorems \ref{theorem1} and \ref{theorem2}, we manage to reduce the channel estimation time duration in Phase III from $(K-1)N$ symbols to
\begin{align}\label{eqn:tau 3 case 2}
\tilde{\tau}_3=\max\left(K-1,\left\lceil \frac{(K-1)N}{M} \right\rceil\right),
\end{align}symbols thanks to the transition from (\ref{eqn:received pilot reflected channel user k 1}) to (\ref{eqn:received pilot reflected channel user k 2}) by exploiting the hidden relation shown in (\ref{eqn:correlated channel}). Further, the designs of user pilot and IRS reflecting coefficients shown in Theorems \ref{theorem1} and \ref{theorem2} are independent of $\mv{g}_{1,n}$'s. Thereby, channel estimation in Phase III does not require any channel feedback from the BS to the users and IRS.

To summarize, to perfectly estimate all the direct channels $\mv{h}_k$'s and reflected channels $\mv{g}_{k,n}$'s in the case without noise at the BS, the minimum pilot sequence length is
\begin{align}\label{eqn:total channel estimation time}
\tilde{\tau}=K+N+\max\left(K-1,\left \lceil\frac{(K-1)N}{M} \right \rceil \right).
\end{align}Interestingly, in the massive MIMO regime \cite{Larsson14}, i.e., $M\rightarrow \infty$, it can be shown that $\tilde{\tau}$ reduces to
\begin{align}\label{eqn:total channel estimation time massive MIMO}
\tilde{\tau}=K+N+K-1=2K+N-1,
\end{align}which is linear with $K$ and $N$. Thereby, under our proposed three-phase channel estimation protocol for IRS-assisted uplink communications, massive MIMO makes it possible to effectively estimate $KMN+KM$ unknown channel coefficients using a scalable number of pilot symbols. Such a result is in sharp contrast to the traditional channel estimation scenario without IRS where the minimum channel estimation time does not depend on the number of receive antennas \cite{Hassibi03}.

\section{Numerical Examples}\label{sec:Numerical Examples}
In this section, we provide numerical examples to verify the effectiveness of our proposed three-phase channel estimation protocol in the IRS-assisted uplink communications. We assume that there are $K=8$ users in the network, and the IRS is equipped with $N=32$ reflecting elements. Moreover, the path loss of $\mv{h}_k$'s, $t_{k,n}$'s, and $\mv{r}_n$'s is modeled as $\beta_k^{{\rm BU}}=\beta_0(d_k^{{\rm BU}}/d_0)^{-\alpha_1}$, $\beta_{k,n}^{{\rm IU}}=\beta_0(d_k^{{\rm IU}}/d_0)^{-\alpha_2}$, and $\beta_n^{{\rm BI}}=\beta_0(d^{{\rm BI}}/d_0)^{-\alpha_3}$, respectively, where $d_0=1$ meter (m) denotes the reference distance, $\beta_0=-20$ dB denotes the path loss at the reference distance, $d_k^{{\rm BU}}$, $d_k^{{\rm IU}}$, and $d^{{\rm BI}}$ denote the distance between the BS and user $k$, between the IRS and user $k$, as well as between the BS and the IRS, while $\alpha_1$, $\alpha_2$, and $\alpha_3$ denote the path loss factors for $\mv{h}_k$'s, $t_{k,n}$'s, and $\mv{r}_n$'s. We set $\alpha_1=4.2$, $\alpha_2=2.1$, and $\alpha_3=2.2$ in the numerical examples. Moreover, the distance between the BS and IRS is set to be $d^{{\rm BI}}=100$ m, and all the users are assumed to be located in a circular regime, whose center is $10$ m away from the IRS and $105$ m away from the BS, and radius is $5$ m. The identical transmit power of users is $23$ dBm.

To illustrate the performance gain of our proposed strategy for channel estimation, we consider a two-phase benchmark strategy. In the first phase, the direct channels $\mv{h}_k$'s are estimated based on (\ref{eqn:perfect channel estimation phase 1}). In the second phase, each user take turns to transmit its pilot sequence such that its reflected channel can be estimated based on (\ref{eqn:perfect channel estimation phase 2}), i.e., (\ref{eqn:correlated channel}) is not exploited.

\begin{figure}[t]
  \centering
  \includegraphics[width=8cm]{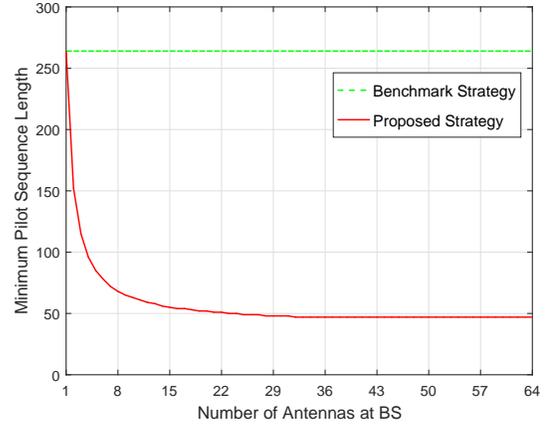}\vspace{-5pt}
  \caption{Minimum pilot sequence length for perfect channel estimation in the case without noise: proposed strategy versus benchmark strategy.}\label{Fig3}\vspace{-10pt}
\end{figure}

Fig. \ref{Fig3} shows the performance comparison between our proposed strategy and the benchmark strategy when the number of antennas at the BS ranges from $M=1$ to $M=64$. It is observed that our proposed strategy can make the best of receive antennas to hugely reduce the channel estimation time. When $M=64$, the minimum pilot sequence length is much shorter compared to the benchmark scheme.

%
\vspace{-2pt}
\section{Conclusion}\label{sec:Conclusion}\vspace{-2pt}
In this paper, we proposed an innovative three-phase protocol to estimate a large number of channel coefficients in the IRS-assisted uplink communications accurately using merely a small number of pilot symbols. Such an interesting result is enabled by exploiting the fact that each IRS elements reflect the signals from all the users to the BS via the same channel. Our results show that time division duplex (TDD) mode is favorable in the future IRS-assisted communication systems because the downlink CSI can be efficiently obtained based on our proposed uplink channel estimation strategy thanks to channel reciprocity.

\begin{appendix}
\subsection{Proof of Theorem \ref{theorem1}}\label{appendix1}
In the case of $M\geq N$, we first prove that there exists a unique solution to (\ref{eqn:received pilot phase 3}) only if $\tau_3 \geq K-1$. Define
\begin{align}\label{eqn:eta}
\eta_{n,i}=\sum\limits_{k=2}^K\lambda_{k,n}\phi_{n,i+\tau_1+\tau_2}a_{k,i+\tau_1+\tau_2}, ~ \forall n, 1\leq i \leq \tau_3.
\end{align}Then, it can be shown that (\ref{eqn:received pilot phase 3}) can be expressed as
\begin{align}\label{eqn:solution 1}
\sum\limits_{n=1}^N\eta_{n,i}\mv{g}_{1,n}= \bar{\mv{y}}^{(\tau_1+\tau_2+i)}, ~~~ i=1,\cdots,\tau_3.
\end{align}Since $t_{k,n}$'s and $\mv{r}_n$'s follow the i.i.d. Rayleigh fading channel model, in the case of $M\geq N$, $\mv{g}_{1,n}$'s are linearly independent with each other with probability one. As a result, $\eta_{n,i}$'s, $n=1,\cdots,N$, $i=1,\cdots,\tau_3$, can be perfectly estimated based on (\ref{eqn:solution 1}). Next, we intend to solve the equations given in (\ref{eqn:eta}). With known $\eta_{n,i}$'s, (\ref{eqn:eta}) characterizes a linear system with $(K-1)N$ variables and $N\tau_3$ equations. Therefore, a unique solution to (\ref{eqn:eta}) exists only when the number of equations is no smaller than the number of variables, i.e., $\tau_3\geq K-1$.

Next, we show that if $\tau_3=K-1$, there always exists a unique solution to (\ref{eqn:received pilot phase 3}) in the case of $M\geq N$. Specifically, if $\tau_3=K-1$, we can set $a_{k,i}$'s and $\phi_{n,i}$'s as given in (\ref{eqn:a case 1}) and (\ref{eqn:b case 1}), respectively. It then follows from (\ref{eqn:received pilot phase 3}) that
\begin{align}\label{eqn:solution 2}
\bar{\mv{y}}^{(\tau_1+\tau_2+k-1)}=[\mv{g}_{1,1},\cdots,\mv{g}_{1,N}]\mv{\lambda}_k, ~~~ k=2,\cdots,K.
\end{align}Since $\mv{g}_{1,n}$'s are linearly independent with each other with probability one in the case $M\geq N$, the pseudo-inverse matrix of $[\mv{g}_{1,1},\cdots,\mv{g}_{1,N}]$ exists. As a result, if $\tau_3=K-1$, there exists a unique solution to (\ref{eqn:received pilot phase 3}), which is given in (\ref{eqn:estimation 1}).

To summarize, in the case of $M\geq N$, there exists a unique solution to (\ref{eqn:received pilot phase 3}) only if $\tau_3 \geq K-1$. Moreover, $\tau_3 =K-1$ is sufficient to guarantee the existence of a unique solution to (\ref{eqn:received pilot phase 3}) by setting $a_{k,i}$'s and $\phi_{n,i}$'s according to (\ref{eqn:a case 1}) and (\ref{eqn:b case 1}). As a result, if $M\geq N$, $\tau_3=K-1$ is the minimum value of $\tau_3$ for perfectly estimating $\mv{\lambda}$ according to (\ref{eqn:received pilot phase 3}).

\subsection{Proof of Theorem \ref{theorem2}}\label{appendix2}
Similar to the case with $M\geq N$, we first prove that in the case of $M<N$, there exists a unique solution to (\ref{eqn:received pilot phase 3}) only if $\tau_3 \geq \lceil \frac{(K-1)N}{M}\rceil$. Note that in (\ref{eqn:received pilot phase 3}), the number of variables and the number of linear equations are $(K-1)N$ and $\tau_3 M$, respectively. As a result, there exists a unique solution to (\ref{eqn:received pilot phase 3}) only if the number of equations is no smaller than that of variables, i.e., $\tau_3 \geq \lceil\frac{(K-1)N}{M}\rceil$.

Next, we show that when $\tau_3 = \lceil\frac{(K-1)N}{M}\rceil$, there always exists a solution to (\ref{eqn:received pilot phase 3}) in the case of $M<N$. The basic idea is that in each time instant $\tau_1+\tau_2+i$ with $i\leq (K-1)\rho$, only one user $k\geq 2$ sends a non-zero pilot symbol for estimating $\lambda_{k,n}$'s with $n\in \Lambda_{k,1}$ without any interference from other users' pilot symbols, while in each time instant $\tau_1+\tau_2+i$ with $(K-1)\rho+1\leq i \leq \tilde{\tau}_3$, multiple users transmit non-zero pilot symbols simultaneously for estimating $\lambda_{k,n}$'s with $n\in \Lambda_{k,2}$'s by eliminating the interference caused by $\lambda_{k,n}$'s with $n\in \Lambda_{k,1}$.

Specifically, at time instant $\tau_1+\tau_2+i$ with $i\leq (K-1)\rho$, we schedule user $k=\lceil \frac{i}{\rho} \rceil+1$ to transmit a pilot symbol $1$, and each of the other users to transmits a pilot symbol $0$, as shown in (\ref{eqn:pilot TDMA 1}). Moreover, only $M$ IRS elements in the set of $\Omega_i$ are switched on and their reflecting coefficients are set to be $1$ as shown in (\ref{eqn:reflecting coefficient TDMA}). In this case, it can be shown that (\ref{eqn:received pilot phase 3}) reduces to
\begin{align}
\bar{\mv{y}}^{(\tau_1+\tau_2+i)} =&\sqrt{p}[\mv{g}_{1,\Omega_i(1)},\hspace{-2pt}\cdots\hspace{-2pt},\mv{g}_{1,\Omega_i(M)}]\nonumber \\ &\times [\lambda_{\lceil\frac{i}{\rho}\rceil+1,\Omega_i(1)},\cdots,\lambda_{\lceil\frac{i}{\rho}\rceil+1,\Omega_i(M)}]^T.
\end{align}Since $\mv{r}_n$'s and $t_{k,n}$'s follow i.i.d. Rayleigh fading channel model, in the case of $M<N$, any $M$ out of $N$ $\mv{g}_{1,n}$'s are linearly independent of each other with probability 1. Therefore, there exists a unique solution to the above equation, which is given by (\ref{eqn:perfect lambda case 1}).

Next, we estimate $\lambda_{k,n}$'s with $n\in \Lambda_{k,2}$. In time instant $\tau_1+\tau_2+i$ with $i \geq (K-1)\rho+1$, all the users in the set $\mathcal{K}_i$ will transmit pilot symbols 1, while each of the other users transmits a pilot symbol $0$, as shown in (\ref{eqn:pilot TDMA 2}). Moreover, all the $M_i\leq M$ IRS elements in the set $\mathcal{N}_i$ are switched on and their reflecting coefficients are set to be $1$ as shown in (\ref{eqn:reflecting coefficient TDMA 2}). In this case, the effective received signal at this time instant given in (\ref{eqn:received pilot reflected channel user k 2}) reduces to
\begin{align}\label{eqn:received signal 2}
\bar{\mv{y}}^{(\tau_1+\tau_2+i)}=\sum\limits_{k\in \mathcal{K}_i}\sum\limits_{n\in \mathcal{N}_i}\sqrt{p}\lambda_{k,n}\mv{g}_{1,n}.
\end{align}Further, for each user $k\in \mathcal{K}_i$, $\lambda_{k,n}$'s with $n\in \Lambda_{k,1}$ have already been perfectly estimated based on (\ref{eqn:perfect lambda case 1}). As a result, their interference can be canceled from (\ref{eqn:received signal 2}) to get $\tilde{\mv{y}}^{(\tau_1+\tau_2+i)}$ shown in (\ref{eqn:signal after interference cancellation}). Moreover, under our construction of $\Lambda_{k,1}$'s and $\Lambda_{k,2}$'s presented prior to Theorem \ref{theorem2}, for any two users $k1,k2\in \mathcal{K}_i$, we have $\Lambda_{k1,2}\cap\Lambda_{k2,2}=\emptyset$ and thus $\Lambda_{k1,2}\subset \Lambda_{k2,1}$ and $\Lambda_{k2,2}\subset \Lambda_{k1,1}$. It can then be shown that $\tilde{\mv{y}}^{(\tau_1+\tau_2+i)}$ reduces to
\begin{align}
\tilde{\mv{y}}^{(\tau_1+\tau_2+i)}\hspace{-2pt}=&\hspace{-2pt} \sqrt{p}[\mv{g}_{1,\mathcal{N}_i(1)},\cdots,\mv{g}_{1,\mathcal{N}_i(M_i)}]\nonumber \\ & \times \left[\hspace{-2pt}\lambda_{\lceil\frac{\mathcal{J}_i(1)}{\upsilon}\rceil+1,\mathcal{N}_i(1)},\cdots,\lambda_{\lceil\frac{\mathcal{J}_i(M)}{\upsilon}\rceil+1,\mathcal{N}_i(M)}\hspace{-2pt}\right]^T,
\end{align}where the $M_i$ elements in the set $\mathcal{N}_i$ can be shown to be different if $\Lambda_{k,2}$'s are constructed based on (\ref{eqn:Lambda 2}). As a result, there exists a unique solution to the above equation which is given by (\ref{eqn:perfect lambda case 2}).

To summarize, in the case of $M<N$, except for the last time instant, we are able to perfectly estimate $M$ unique $\lambda_{k,n}$'s either via (\ref{eqn:perfect lambda case 1}) or (\ref{eqn:perfect lambda case 2}) at each time instant, while at the last time instant, the remaining $\lambda_{k,n}$'s are estimated. As a result, the minimum $\tau_3$ for perfect channel estimation is characterized by (\ref{eqn:tau 3 case 2}). Theorem \ref{theorem2} is thus proved.

\end{appendix}

\bibliographystyle{IEEEtran}
\bibliography{CIC}

\begin{thebibliography}{10}
\providecommand{\url}[1]{#1}
\csname url@samestyle\endcsname
\providecommand{\newblock}{\relax}
\providecommand{\bibinfo}[2]{#2}
\providecommand{\BIBentrySTDinterwordspacing}{\spaceskip=0pt\relax}
\providecommand{\BIBentryALTinterwordstretchfactor}{4}
\providecommand{\BIBentryALTinterwordspacing}{\spaceskip=\fontdimen2\font plus
\BIBentryALTinterwordstretchfactor\fontdimen3\font minus
  \fontdimen4\font\relax}
\providecommand{\BIBforeignlanguage}[2]{{%
\expandafter\ifx\csname l@#1\endcsname\relax
\typeout{** WARNING: IEEEtran.bst: No hyphenation pattern has been}%
\typeout{** loaded for the language `#1'. Using the pattern for}%
\typeout{** the default language instead.}%
\else
\language=\csname l@#1\endcsname
\fi
#2}}
\providecommand{\BIBdecl}{\relax}
\BIBdecl

\bibitem{Liaskos08}
C.~Liaskos, S.~Nie, A.~Tsioliaridou, A.~Pitsillides, S.~Ioannidis, and
  I.~Akyildiz, ``{A new wireless communication paradigm through
  software-controlled metasurfaces},'' \emph{IEEE Commun. Mag.}, vol.~56,
  no.~9, pp. 162--169, Sep. 2018.

\bibitem{Renzo19}
M.~D.~R. {\it et al.}, ``{Smart radio environments empowered by reconfigurable
  AI meta-surfaces: An idea whose time has come},'' \emph{EURASIP J. Wireless
  Commun. Network.}, no. 129, pp. 1--20, May 2019.

\bibitem{Basar19}
E.~Basar, M.~D. Renzo, J.~Rosny, M.~Debbah, M.-S. Alouini, and R.~Zhang,
  ``{Wireless communications through reconfigurable intelligent surfaces},''
  \emph{IEEE Access}, vol.~7, pp. 116\,753--116\,773, 2019.

\bibitem{Wu18}
Q.~Wu and R.~Zhang, ``{Intelligent reflecting surface enhanced wireless
  network: joint active and passive beamforming design},'' in \emph{Proc. IEEE
  Global Commun. Conf. (Globecom)}, Dec. 2018.

\bibitem{Zhang19}
S.~Zhang and R.~Zhang, ``{Capacity characterization for intelligent reflecting
  surface aided MIMO communication},'' [Online] Available:
  \url{https://arxiv.org/abs/1910.01573}.

\bibitem{Larsson14}
E.~G. Larsson, F.~Tufvesson, O.~Edfors, and T.~L. Marzetta, ``{Massive MIMO for
  next generation wireless systems},'' \emph{IEEE Commun. Mag.}, vol.~52,
  no.~2, pp. 186--195, Feb. 2014.

\bibitem{Liu18}
L.~Liu and W.~Yu, ``{Massive connectivity with massive MIMO-Part I: Device
  activity detection and channel estimation},'' \emph{IEEE Trans. Signal
  Process.}, vol.~66, no.~11, pp. 2933--2946, June 2018.

\bibitem{Mishra19}
D.~Mishra and H.~Johansson, ``{Channel estimation and low-complexity
  beamforming design for passive intelligent surfaceassisted MISO wireless
  energy transfer},'' in \emph{Proc. IEEE Int. Conf. Acoustics Speech Signal
  Process. (ICASSP)}, May 2019.

\bibitem{Yang19}
Y.~Yang, B.~Zheng, S.~Zhang, and R.~Zhang, ``{Intelligent reflecting surface
  meets OFDM: Protocol design and rate maximization},'' [Online] Available:
  \url{https://arxiv.org/abs/1906.09956}.

\bibitem{Zheng19}
B.~Zheng and R.~Zhang, ``{Intelligent reflecting surface-enhanced OFDM: Channel
  estimation and reflection optimization},'' [Online] Available:
  \url{https://arxiv.org/abs/1909.03272}.

\bibitem{Jensen}
T.~L. Jensen and E.~D. Carvalho, ``{On optimal channel estimation scheme for
  intelligent reflecting surfaces based on a minimum variance unbiased
  estimator},'' [Online] Available: \url{https://arxiv.org/abs/1909.09440}.

\bibitem{Hassibi03}
B.~Hassibi and B.~M. Hochwald, ``{How much training is needed in
  multiple-antenna wireless links},'' \emph{IEEE Trans. Inf. Theory}, vol.~49,
  no.~4, pp. 951--963, Apr. 2013.

\end{thebibliography}

\end{document}